\documentclass[twocolumn,english,amssymb,prd,nofootinbib]{revtex4-2}
\usepackage{amsfonts,bm,amssymb,euscript,array,babel}
\usepackage{mathtools}
\usepackage{tikz-cd}
\usepackage{amsmath}
\usepackage{hhline}
\usepackage[amsmath]{empheq}
\usepackage{hyperref}
\usepackage{cancel}
\usepackage{mathrsfs}
\usepackage{lipsum}
\usepackage{adjustbox}

\newcommand{\dd}{\mathrm{d}}

\newcommand{\be}{\begin{equation}}
\newcommand{\ee}{\end{equation}}

\def\nn{\nonumber}
\def \bea{\begin{eqnarray}} 
\def\eea{\end{eqnarray}}
\def\bse{\begin{subequations}}	
	\def\ese{\end{subequations}}
\def\bal{\begin{align}} 
\def\eal{\end{align}}

\def\bi{\begin{itemize}} 
	\def\ei{\end{itemize}}

\def\a{\alpha} \def\b{\beta}    
\def\e{\epsilon} 
  \def\h{\eta} 
  
 \def\o{\omega}

\def\one{\mbox{1 \kern-.59em {\rm l}}}

\numberwithin{equation}{section}

\begin{document}

	\title{Relation between standard and exotic duals of differential forms}
		
	\author{Athanasios Chatzistavrakidis}
	\email{Athanasios.Chatzistavrakidis@irb.hr}
	\author{Georgios Karagiannis}
	\email{Georgios.Karagiannis@irb.hr}
		\affiliation{Division of Theoretical Physics, Rudjer Bo\v skovi\'c Institute, Bijeni\v cka 54, 10000 Zagreb, Croatia }
		\begin{abstract}
			Exotic duality suggests a link between gauge theories for differential $p$-forms and tensor fields of mixed symmetry $[D-2,p]$ in $D$ spacetime dimensions. On the other hand, standard Hodge duality relates $p$-form to $(D-p-2)$-form gauge potentials by exchanging their field equations and Bianchi identities. Following the methodology and the recent proposal of  Henneaux \emph{et al.} that the double dual of the free graviton is algebraically related to the original graviton and does not provide a new, independent description of the gravitational field, we examine the status of exotic duality for $p$-forms. We find that the exotic dual is algebraically related to the standard dual of a differential form and therefore they provide equivalent descriptions as free fields. Introducing sources then leads to currents being proportional. This relation is extended in a straightforward way for higher exotic duals of the mixed symmetry type $[D-2,\dots,D-2,p]$.  
		\end{abstract}
	\maketitle
\section{Introduction}
Maxwell's equations in vacuum, or in the presence of both electric and magnetic charges, exhibit a duality in the sense that they may be described in terms of two different gauge potentials. Electromagnetic duality exchanges the field equations of one field with the Bianchi identities of the other and vice versa. 

A similar duality holds in linearized gravity, where the graviton can be dualized to a field of mixed symmetry type $[D-3,1]$ in $D$ dimensions \cite{Hull:2001iu,West:2001as}. For instance, in five dimensions it is dual to a $[2,1]$ $GL(D)$-irreducible tensor field \cite{Boulanger:2003vs}, whose gauge theory was described by Curtright \cite{Curtright:1980yk}. In addition, it was suggested in \cite{Hull:2001iu} that the graviton possesses a double dual field of type $[D-3,D-3]$. However, it was noted that despite the apparent existence of three mutually dual fields, only two sources arise \cite{Hull:2001iu}. 

In Ref. \cite{Henneaux:2019zod}, it was shown that the double dual graviton in $D$ spacetime dimensions is algebraically related to the standard graviton $[1,1]$. For instance in $D=5$ dimensions, considering the three candidate mutually dual fields 
$ 
h_{[1,1]}, C_{[2,1]}, \hat{h}_{[2,2]}\,,
$ 
one may ask how many of them are not algebraically related.
The answer in \cite{Henneaux:2019zod} is two, 
which may be depicted by the following diagram
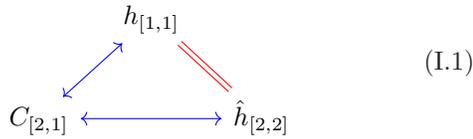
\be \label{gravitondiagram}\begin{tikzcd}[column sep=small]
	& h_{[1,1]}\arrow[dl,leftrightarrow,blue] \arrow[equal,red,dr] & \\
	C_{[2,1]} \arrow[leftrightarrow,blue]{rr} & & \hat{h}_{[2,2]}
\end{tikzcd}
\ee
where blue arrows indicate standard Hodge duality of the corresponding field strengths and the red double line indicates the aforementioned algebraic equivalence. The conclusion is that the role of the double dual is different than that of the standard dual field, and its status is not of one providing an independent description of the gravitational field, in the sense that field equations and Bianchi identities are not exchanged in that case. This provides additional evidence for the absence of a doubly magnetic source at the level of linearized gravity.

Motivated by the above, in this letter we focus instead on a differential $p$-form in $D$ dimensions. It has been argued that apart from the standard $(D-p-2)$-form dual, there exists an infinite  number of ``exotic'' duals \cite{Boulanger:2015mka}. These are irreducible mixed-symmetry tensor fields of types $[D-2,\dots,D-2,p]$ and $[D-2,\dots,D-2,D-p-2]$, essentially obtained by appending columns of size $D-2$ in the corresponding Young tableaux. For example, Maxwell theory has an alternative description in terms of a $[2,1]$ field, and similar examples exist for higher forms \cite{Boulanger:2015mka,Bergshoeff:2016ncb,Bergshoeff:2016gub}. The natural question is then how many out of these infinite dual fields are truly independent?

We address the above question using the methodology of \cite{Henneaux:2019zod}. In Sec. \ref{Section 2}, we begin by truncating the analysis at the level of bipartite tensor fields, namely ones corresponding to a Young tableau with two columns. We proceed to show that out of the four possible mutually dual descriptions of a free spin-1 $p$-form only two are truly independent. In Section \ref{Section 2C} this result is extended to the infinite chains of dualities mentioned above. In Section \ref{Section 3} we comment on the implications of this result for sources and we collect our conclusions and further comments in Sec. \ref{Section 4}.

\section{Standard vs. Exotic Duality}\label{Section 2}

In this section, we start with dual fields having up to two
sets of antisymmetrized indices, namely, with symmetry
represented by Young tableaux with up to two columns. We
refer to them as bipartite tensor fields. Later in this section,
we will also encounter dual mixed symmetry tensor fields
with more sets of antisymmetrized indices, namely, multipartite tensor fields. All such fields can be elegantly
described in a formalism that generalizes differential forms,
as follows. We introduce $N$ sets of anticommuting variables
$\theta^{\mu}_\a$, $\a=1,\dots,N$ satisfying
\be \label{2.1}
\theta^{\mu}_\a\theta^{\nu}_\a=-\theta^{\nu}_\a\theta^{\mu}_\a\,,\qquad \theta^{\mu}_\a\theta^{\nu}_{\b\neq\a}=\theta^{\nu}_{\b\neq\a}\theta^{\mu}_\a\,;
\ee
namely, the variables of each set are anticommuting, but
they commute with the ones of all other sets. Associating
them to odd coordinates on a (graded) manifold of which
the bosonic base is equipped with (even) coordinates $x^\mu$,
any $N$-partite tensor field $A$ becomes a function that is
expanded in the above basis as
\be \label{2.2}
A=\frac{A_{[\mu_1^1\dots\mu^1_{p_1}]\dots[\mu_1^N\dots\mu^N_{p_N}]}(x)}{\prod_{k=1}^Np_k!}\,\theta_1^{\mu^1_1}\dots\theta_1^{\mu^1_{p_1}}\dots\theta_N^{\mu^N_1}\dots\theta_N^{\mu^N_{p_N}},
\ee
with the desired index symmetry for its components
inherited automatically by \eqref{2.1}. We then say that the
tensor has multidegree $[p_1,\dots,p_N]$.

This observation allows us to develop a differential
geometry analogous to $p$-forms, which correspond to the
case $N=1$. In particular, we define $N$ exterior derivatives
$\dd_\a=\theta^\mu_\a\partial_\mu$, which raise by 1 the $\a$th slot of the multidegree
of $A$ and satisfy $\dd_\a^2=0$ and $\dd_\a\dd_\b=\dd_\b\dd_\a$ for $\a\neq\b$. For
example, when $N=2$ and $p_1=p_2=1$, the exterior
derivative $\dd_1$ acts as
\be 
\dd_1A=\partial_{[\mu}A_{\nu]\rho}\theta_1^{\mu}\theta_1^\nu\theta_2^\rho,
\ee
thus yielding a bipartite tensor of bidegree $[2,1]$. Moreover,
$N$ mutually commuting Hodge star operators $\ast_\a$ are defined
in complete analogy to $p$-forms, using the Levi-Civita
symbol to dualize in the standard way the $\a$th slot of the
multidegree of $A$ from $p_\a$ to $D-p_\a$ in $D$ dimensions.

An additional ingredient in the differential geometry of
multipartite tensors, which is atypical for $p$-forms, is a set
of $(N-1)!$ transposition maps $\top_{\a\b}$ that act on $A$ by simply
interchanging $\theta_\a^\mu\leftrightarrow\theta_\b^\mu$ in \eqref{2.2}. Unlike $p$-forms, multipartite tensors have traces $\text{tr}_{\a\b}$ in general, which may be
defined simply replacing $\theta_\a^{\mu^\a_1}\theta_\b^{\mu^\b_1}\leftrightarrow p_\a p_\b \h^{\mu_1^\a\mu_1^\b}$ in \eqref{2.2},
where $\h^{\mu\nu}$ is the inverse Minkowski metric. We note in
passing that one may encode the components of the
Minkowski metric in a $[1,1]$ bipartite tensor, e.g. $\h=\h_{(\mu\nu)}\theta_1^\mu\theta_2^\nu$,  when $N=2$. These are then all the basic
ingredients we will use in what ensues. Further details
on this formalism may be found in Refs. \cite{deMedeiros:2002qpr,ChKS}.

Our notation for dual fields is the following. We begin
with a differential $p$-form $A^{(0)}$ in $D$ dimensions. The
superscript in parentheses refers to the level, which will
be more transparent below. We refrain from explicitly
denoting the degree of the tensor as a subscript whenever
it is clear from context and reinstate it when necessary. The
standard $(D-p-2)$-form dual will be denoted by $B^{(0)}$.
Exotic dual tensor fields of types $[D-2,\dots,D-2,p]$ and
$[D-2,\dots,D-2,D-p-2]$, where in each case we have
$n\geq 1$ columns of size $D-2$, will be denoted by either $A^{(n)}$
or $B^{(n)}$ depending on whether $n$ is even or odd. The fields
$A^{(\text{even})}$ and $B^{(\text{odd})}$ will be of type $[D-2,\dots,D-2,p]$,
while $A^{(\text{odd})}$ and $B^{(\text{even})}$ will correspond to $[D-2,\dots,D-2,D-p-2]$ Young tableaux.
\subsection{\label{SHD} Non-standard approach to standard  duality}\label{Section 2A}

Presently, we focus on bipartite tensors, and the main
question we would like to address becomes how many out
of the four mutually dual fields $A^{(0)}_p$, $B^{(0)}_{D-p-2}$, $A^{(1)}_{[D-2,D-p-2]}$ and $B^{(1)}_{[D-2,p]}$ are not algebraically related. As  already mentioned, the fields $A^{(0)}$ and $B^{(0)}$ are related by standard electromagnetic duality. Rather than presenting the well-known way to show this, for later purposes we follow here a somewhat unorthodox but completely equivalent route. 
The main trick is to think of $A^{(0)}$ as a (degenerate) bipartite tensor of type $[p,0]$ and define the irreducible $[p+1,1]$ Riemann-like tensor
\be\label{curv B2}
R^{A^{(0)}}:=\dd_1\dd_2A^{(0)}\,.
\ee
Because of the nilpotency of the two exterior derivatives, the tensor $R^{A^{(0)}}$ satisfies by definition the Bianchi identities
\be\label{BI RB}
\dd_1 R^{A^{(n)}}=0=\dd_2 R^{A^{(n)}}\,,
\ee
for $n=0$.
On the other hand, the field equations for the free differential form $A^{(0)}$ are simply given by the free wave equation
 \be \label{eoms A0}\dd_1\ast_1\dd_1\,A^{(0)}=0\,.\ee 
Evidently, the subscripts 1 may be simply ignored for a $p$-form; however,  the interesting point is that this field equation can be equivalently written as
\be\label{EOMS B2}
\text{tr} R^{A^{(0)}}=0\,,
\ee 
where $\text{tr}\equiv\text{tr}_{12}$. This is true due to the identity \cite{deMedeiros:2002qpr}
\be \label{mapidentity1}
\dd^{\dagger}_1=\dd_2\,\text{tr}+\text{tr}\,\dd_2\,,
\ee 
where $\dd^{\dagger}_1:=(-1)^{1+D(p+1)} \ast_1\dd_1\,\ast_1$ is the co-differential of $\dd_1$, which reduces the degree of the first slot of a bipartite tensor by 1---being essentially the divergence. 

Then, one can define another irreducible bipartite tensor of type $[D-p-1,1]$ by means of the operator $\ast_1$,
\be\label{RC}
R^{B^{(0)}}:=\ast_1R^{A^{(0)}}\,.
\ee
Its irreducibility is implied by the field equations \eqref{EOMS B2} of $A^{(0)}$.
Using the Bianchi identities \eqref{BI RB} and the field equation \eqref{EOMS B2} one can easily show that $R^{B^{(0)}}$ also satisfies the corresponding  Bianchi identities, i.e. 
\be\label{BI RC}
\dd_1 R^{B^{(n)}}=0=\dd_2 R^{B^{(n)}}\,, 
\ee
for $n=0$.
Because of that, one can locally identify $R^{B^{(0)}}$ as the Riemann tensor for the $(D-p-2)$-form standard dual field $B^{(0)}$, namely
\be\label{C6}
R^{B^{(n)}}:=\dd_1\dd_2B^{(n)}
\ee
for $n=0$. Then the field equations for $B^{(0)}$ follow trivially from the definition \eqref{RC}. The tensor $R^{B^{(0)}}$ is identically traceless, since $R^{A^{(0)}}$ is irreducible, so we obtain the field equations
\be\label{field equations C}
\text{tr} R^{B^{(0)}}=0\,,
\ee
which by the same reasoning as before are just the free wave equation for the dual field.
Finally, the relation between $A^{(0)}$ and $B^{(0)}$ can be directly read off from \eqref{RC} and it is
\be\label{relation BC}
\dd_1 B^{(0)}=\ast_1\,\dd_1 A^{(0)}\,,
\ee
up to a physically irrelevant constant $(D-p-1)$-form. Therefore, the relation among the two fields is 
identical to the one obtained by standard method and it is non-local, which means that $A^{(0)}$ and $B^{(0)}$ cannot be algebraically related. 

\subsection{\label{AE} Equivalence of standard \& exotic duals}\label{Section 2B}

Exotic duality in the above spirit may be understood as the Hodge duality of the original $p$-form field along its second, trivial slot. In other words, one would replace the Hodge star $\ast_1$ in \eqref{RC} by the second Hodge star $\ast_2$ and define 
\be \label{definition RB1}
\left(R^{B^{(1)}}\right)^{\top}=\ast_2\,R^{A^{(0)}}\,,
\ee
where $\top\equiv\top_{12}$ is the transposition map defined in the beginning of this section. The tensor $R^{B^{(1)}}$ is then of type $[D-1,p+1]$, it is irreducible and it satisfies the Bianchi identities \eqref{BI RC} for $n=1$. Locally, it is given by \eqref{C6} for $n=1$ and the respective gauge field $B^{(1)}$ satisfies the field equations 
\be\label{equations B1}
\text{tr}^{\,p+1}R^{B^{(1)}}=0\,,
\ee
which follow identically from the definition \eqref{definition RB1}.
Then, the same reasoning as before leads to the duality relation
\be \label{relation B1 A0}
\dd_2(B^{(1)})^{\top}=\ast_2\,\dd_2\,A^{(0)}\,,
\ee  
up to a bi-closed $[p,D-1]$ diffeomorphism. 

Up to this point it appears that $B^{(1)}$ is another, inequivalent dual field. However, one may now ask what is its relation to $B^{(0)}$. 
According to the definitions above and the simple fact that the Hodge star squares to unity up to a sign when it acts on a $p$-form, $(\ast_1)^2=(-1)^{1+p(D-p)}$, we immediately find that
\be \label{RB1 RB0}
\left(R^{B^{(1)}}\right)^{\top}=(-1)^{1+(p+1)(D-p-1)}\ast_1\ast_2\,R^{B^{(0)}}\,.
\ee 
Taking into account the field equation \eqref{field equations C}, one can show that the above relation implies that
\be\label{RB1 ETA}
R^{B^{(1)}}\propto\,\eta^{\,p}R^{B^{(0)}}\,.
\ee 
Here, $\eta$ is the $[1,1]$ tensor field whose components are those of the Minkowski metric and the relation holds up to an irrelevant constant prefactor that can be absorbed into the definition of $B^{(0)}$. Eq. \eqref{RB1 ETA} is easily proven by considering a ``full'' Hodge star operator $\star$ acting on a (not necessarily irreducible) $(p,q)$ bipartite tensor field $\o$ and returning a $(D-p,D-q)$ one, 
\be\label{FullHodge}
(\star\,\o)_{D-p,D-q}=\frac{1}{(D-p-q)!}\,\eta^{D-p-q}\,{\o}^{\top}_{q,p}\,,\ee
first defined in \cite{Chatzistavrakidis1} and further explained in \cite{ChKS}. 
This is a operation distinct from the product of the partial Hodge stars; specifically, it satisfies 
	\be \label{fullhodge2}
\star \o=(-1)^{\e}\ast_1\ast_2\, \sum_{n=0}^{\text{min}(p,q)}\frac {(-1)^n}{(n!)^2}\,\eta^n\,\text{tr}^n\,\o~,\ee
where $\e=(D-1)(p+q)+pq+1$. Applying this formula for the $[D-p-1,1]$ tensor $R^{B^{(0)}}$ and using \eqref{FullHodge} along with \eqref{field equations C}, the desired relation follows. 

Finally, using the local expressions for these Riemann tensors and the Poincar\'e lemma, one ends up with the relation
\be \label{algebr B1 B0}
B^{(1)}\propto \eta^{\,p}B^{(0)}\,,
\ee
which proves that $B^{(1)}$ is algebraically related to $B^{(0)}$ up to a $\dd_1\dd_2$-closed $[D-2,p]$-type diffeomorphism.

Equivalently, one can define another tensor by 
\be \label{definition RA1}
\left(R^{A^{(1)}}\right)^{\top}=\ast_2\,R^{B^{(0)}}=\ast_1\ast_2R^{A^{(0)}}
\ee
and follow the same procedure. This leads to additional duality and algebraic relations, specifically
\begin{align}
&\dd_2(A^{(1)})^{\top}=\ast_2\,\dd_2\,B^{(0)}\,,\label{relations A1 B0}\\
&A^{(1)}\propto\eta^{D-p-2}A^{(0)}\,,\label{relation A1 A0}
\end{align}
up to diffeomorphisms of suitable degree.
The last relation that we would like to mention reads as
\be\label{relation A1 B1}
\dd_2A^{(1)}\propto \ast_2\,\dd_2\,B^{(1)}\,,
\ee
and follows from \eqref{RC},  \eqref{definition RB1} and \eqref{definition RA1}. 

Thus, we have found the six mutual relations among the four fields under investigation, which are \eqref{relation BC}, \eqref{relation B1 A0}, \eqref{relations A1 B0}, \eqref{algebr B1 B0}, \eqref{relation A1 A0} and \eqref{relation A1 B1}. The first and last of these relations show the standard Hodge duality relations between $A^{(0)}$-$B^{(0)}$ and $A^{(1)}$-$B^{(1)}$, respectively. The second and third relations, which are actually what is known as exotic duality relations, can be also seen to be standard Hodge duality relations. In fact, they imply that exotic duality is just standard Hodge duality of the transposed fields $(A^{(0)})^\top$ and $(B^{(0)})^\top$.
Finally, the fourth and fifth are \emph{not} duality relations and show that the exotic dual fields $A^{(1)}$ and $B^{(1)}$ are algebraically related to $A^{(0)}$ and $B^{(0)}$ respectively.

We conclude with a diagram analogous to \eqref{gravitondiagram} depicting these results, valid for any spacetime dimension $D$:
\be \label{diagram first level}
\adjustbox{scale=0.85,center}
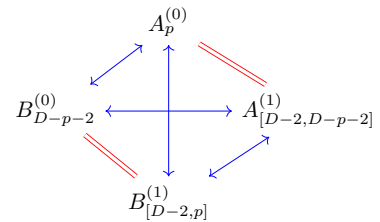
{\begin{tikzcd}[column sep=tiny]
             & A^{(0)}_p \arrow[dl,leftrightarrow,blue] \arrow[dr,equal,red] \arrow[dd,blue,leftrightarrow]& \\
B^{(0)}_{D-p-2} \arrow[rr,blue, leftrightarrow] \arrow[dr,equal,red] &                         & A^{(1)}_{[D-2,D-p-2]}\arrow[dl,leftrightarrow,blue]\\
             & B^{(1)}_{[D-2,p]}
\end{tikzcd}}\nn
\ee 
One may interpret the red double lines as double duality in the same way as for the graviton. This is in full agreement with the analogous result in \cite{Henneaux:2019zod} that double duals are algebraically related to the original field.  Alternatively, one may think of all closed loops of this diagram as commutative; for instance, exotic duality is then just standard duality followed by an equivalence (double duality).

Interesting examples include the 1-form in four and the 2-form in ten dimensions. In the former case, one deals with standard Maxwell theory, where the electric and magnetic dual potentials are both 1-forms $A^{(0)}_{\mu}$ and ${B}^{(0)}_{\mu}$. Following \cite{Boulanger:2015mka}, there is yet another way to describe the theory in terms of a $[2,1]$ gauge potential $B^{(1)}_{[\mu\nu]\rho}$. This is indeed a dual of the original potential $A^{(0)}_{\mu}$. However, physically it yields the same description as the $B^{(0)}_{\mu}$ potential in the sense that there is no new source other than a magnetic monopole associated to it, no additional global symmetries apart from the standard electric and magnetic ones $U(1)_{e}\times U(1)_{m}$ and no new loop operator aside the standard Wilson and 't Hooft loops. 

Similarly, the 2-form in ten dimensions, which may be identified with the Kalb-Ramond field of the common sector of closed string theories, dualizes to a 6-form and an $[8,2]$ mixed symmetry tensor field respectively. Our findings indicate that these two duals are equivalent in the sense described above. We comment on the implications for the corresponding sources below.

\subsection{Higher exotic duals}\label{Section 2C}

We now turn to the more general case of exotic duality for differential forms, which refers to multipartite tensor fields of types $[D-2,\dots,D-2,p]$ and $[D-2,\dots, D-2,D-p-2]$ \cite{Riccioni:2006az,Boulanger:2015mka}. Their relation to the original $p$-form field and its standard dual may be found in a way similar to Sec. \ref{Section 2B}.

In particular, one can advance in the next level by interpreting a $p$-form as a $[p,0,0]$ tripartite tensor and then defining the higher $[D-1,D-1,p+1]$-type  Riemann tensor $R^{A^{(2)}}$ as
\be 
\left(R^{A^{(2)}}\right)^{\top_{13}}:=\ast_2\ast_3\dd_3R^{A^{(0)}}\label{RA2},
\ee
similarly to what we did in \eqref{definition RA1}.
$R^{A^{(2)}}$ is irreducible and satisfies all three Bianchi identities, i.e. it is closed with respect to $\dd_\a$ for any $\a=1,2,3$.

Locally, the dual field $A^{(2)}$ can be introduced as 
\be\label{local expression RA2}
R^{A^{(2)}}:=\dd_1\,\dd_2\,\dd_3\,A^{(2)}\,,
\ee
the field equations, following from \eqref{RA2}, of which are
\be
\text{tr}_{12}^{D-1}R^{A^{(2)}}=0=\text{tr}_{23}^{\,p+1}R^{A^{(2)}}.
\ee
Following the analogous steps as in the previous section, Eq. \eqref{RA2} leads to the relation
\be 
A^{(2)}\propto \,\eta_{12}^{\,D-2}\,(A^{(0)})^{\top_{13}},
\ee
up to a $\dd_1\dd_2\dd_3$-closed $[D-2,D-2,p]$ tensor. We observe once more that this is an algebraic relation.
Moreover, a similar result holds for the dual field $B^{(2)}$, which has a Riemann tensor defined as in \eqref{RA2} but with $A\to B$. It turns out to be algebraically related to the standard dual $B^{(0)}$.

Overall, the procedure described above can be repeated as many times desired for higher levels. This results in two separate infinite chains of algebraic equivalences; one corresponding to $A^{(0)}$ and the other to $B^{(0)}$. 
The  diagram depicting the aforementioned chains of dualities and algebraic equivalences is
	\be\begin{tikzcd}
		A^{(0)} \arrow[dr,blue,leftrightarrow]\arrow[d,blue,leftrightarrow]\arrow[r,equal,red]
		& A^{(1)} \arrow[dr,blue,leftrightarrow]\arrow[r,equal,red]\arrow[d,blue,leftrightarrow]
		& \cdots\arrow[dr,blue,leftrightarrow]\arrow[r,equal,red]\arrow[d,blue,leftrightarrow] & A^{(n)} \arrow[dr,blue,leftrightarrow]\arrow[r,equal,red]\arrow[d,blue,leftrightarrow]
		& \cdots \\
		B^{(0)} \arrow[ur,blue,leftrightarrow]\arrow[r,equal,red]
		& B^{(1)} \arrow[ur,blue,leftrightarrow]\arrow[r,equal,red]
		& \cdots\arrow[ur,blue,leftrightarrow]\arrow[r,equal,red] & B^{(n)} \arrow[ur,blue,leftrightarrow]\arrow[r,equal,red]
		&\cdots
	\end{tikzcd}\nn\ee
As before, the vertical (diagonal) blue arrows indicate standard (exotic) Hodge duality, while the algebraic equivalences are depicted by the red double lines.

\section{Comments on sources}\label{Section 3}

In theories exhibiting electric/magnetic duality, one can correspondingly introduce two currents coupling to the original and dual gauge potentials. 
In the non-standard approach to $p$-form gauge theory of Sec. \ref{Section 2A}, 
the electric $p$-form current $j^{A^{(0)}}$ is introduced as 
\be\label{source RB}
\text{tr}R^{A^{(0)}}=j^{A^{(0)}}\,,
\ee 
and the Bianchi identities \eqref{BI RB} imply its conservation, namely $\dd^\dagger_1 j^{A^{(0)}}=0$. This is shown using the identity \cite{deMedeiros:2002qpr} 
\be \dd^\dagger_1\text{tr}=\frac{1}{2}\dd_2\,\text{tr}^2-\frac{1}{2}\text{tr}^2\,\dd_2\,.\ee

In the presence of the electric current, the dual tensor $R^{B^{(0)}}$ defined in  \eqref{RC} is no longer irreducible. Indeed, the deformed field equations \eqref{source RB} imply that $\ast_1\text{tr}\ast_1 R^{B^{(0)}}\propto \ast_1 j^{A^{(0)}}\neq 0$, which obstructs its irreducibility according to the criterion of \cite{deMedeiros:2002qpr}. In addition, this reducible tensor does not obey both Bianchi identities, since 
\be\label{mod BI RC}
\dd_1 R^{B^{(0)}}\propto \ast_1\,\dd_2j^{A^{(0)}}\ne 0\,,\quad \dd_2R^{B^{(0)}}=0\,.
\ee
Thus, the Poincar\'e lemma cannot be used to introduce the dual field $B^{(0)}$ in that case.

Alternatively, one could start with the 
irreducible $[D-p-1,1]$-type Riemann 
tensor $R^{B^{(0)}}:=\dd_1\dd_2B^{(0)}$, which obeys the Bianchi identities \eqref{BI RC}, and introduce a magnetic $(D-p-2)$-form current $j^{B^{(0)}}$ as
\be\label{source RC}
\text{tr}R^{B^{(0)}}=j^{B^{(0)}}\,.
\ee
Then, by the same logic, the potential $A^{(0)}$ for $R^{A^{(0)}}$ defined as in \eqref{RC} cannot be introduced.

 Let us now turn to sources associated to the exotic dual field $A^{(1)}$.
Suppose that we are again in a spacetime region where an electric current is present, i.e. $j^{A^{(0)}}\neq 0$. As already discussed, the theory can be described by the gauge potential $A^{(0)}$ satisfying the field equation \eqref{source RB}, but \emph{not} with $B^{(0)}$. What about $A^{(1)}$? 
Note that unlike $R^{B^{(0)}}$, the Riemann tensor $R^{A^{(1)}}$ is still irreducible. However, it does not satisfy the Bianchi identities, since
\be\label{BI new RD}\begin{split}
	&\dd_1 R^{A^{(1)}}\propto \ast_1\ast_2\,\dd_2(j^{A^{(0)}})^{\top}\neq 0\\
	&\dd_2 R^{A^{(1)}}\propto \ast_1\ast_2\,\dd_1(j^{A^{(0)}})^{\top}\neq 0\,.
\end{split}
\ee
An ``exotic'' $p$-form current $j^{A^{(1)}}$ would be introduced then as
\be\label{source RD}
\text{tr}^{\,D-p-1}R^{A^{(1)}}=j^{A^{(1)}}\,.
\ee
However, one can now easily see that the two currents are proportional 
\be\label{sources RB RD}
j^{A^{(1)}}\propto j^{A^{(0)}}\,,
\ee
which is to be expected by means of our analysis in the previous sections. The same approach shows that the would-be exotic current for $B^{(1)}$ is proportional to the one for $B^{(0)}$. This is in complete analogy to the graviton case, where the current associated to its double dual is proportional to the original energy-momentum tensor \cite{Hull:2001iu}.

\section{Conclusions} \label{Section 4}

In this letter, we studied the relation between standard and exotic dual fields of differential forms. Motivated by the analysis and conclusions of \cite{Henneaux:2019zod} regarding the double dual of the free graviton and its status as a true dual field, we investigated whether there are algebraic relations among different dual descriptions of a free $p$-form gauge theory. Our results indicate that only two dual fields should be thought of as providing inequivalent descriptions of the theory, the rest being connected to them by two (infinite) towers of  algebraic relations. This is also true for the associated currents that can couple to the different gauge potentials. Thus, for example one may choose to consider as dual of a $p$-form either a $(D-p-2)$-form or a $[D-2,p]$ mixed symmetry tensor with certain properties, but these lead to the same dual description of the theory on-shell \footnote{Nevertheless, off-shell parent actions may differ in their field content, including auxiliary fields; see for example \cite{Boulanger:2012df,Boulanger:2015mka,Bergshoeff:2016ncb,ChKS}.}.

Here we have considered only free and single fields. It would be interesting to think about the implications of our result for different dual descriptions of string theory in terms of mixed symmetry tensors and for the Wess-Zumino couplings of exotic branes \cite{deBoer:2012ma}. Regarding the latter, it has been argued group theoretically \cite{West:2004kb,Bergshoeff:2010xc,Bergshoeff:2011se} and using T- and S-duality \cite{Chatzistavrakidis:2013jqa,Kimura:2014upa} that they couple to mixed symmetry tensor fields, which in turn appear naturally in the $E_{11}$ approach to M-theory \cite{West:2001as,Riccioni:2006az}. String and M-theory contain a host of different fields and their brane couplings include non-linear contributions. 
In addition, applying T-duality on the NS5 brane couplings can be tricky due to world-sheet instanton corrections that can break isometries \cite{Tong:2002rq} (see also \cite{Becker:2009df}, and \cite{Kimura:2013zva} for exotic branes.) Therefore, the results of the present paper cannot be applied as such to questions regarding brane couplings, however they indicate that caution should be exercised regarding the definition and properties of exotic branes. 

Finally, sources for bipartite tensor fields were considered  in Ref. \cite{Bunster:2013era}. It was shown that the extended object coupling to such a field is a combined one, comprising two branes of different dimensionality whose world-volumes interweave. It would be interesting to revisit this construction under the light of the relations proposed here.

\begin{acknowledgements}
	The authors would like to thank M. Henneaux, V. Lekeu and A. Leonard for  useful discussions. 
		This work is supported by the Croatian Science Foundation Project ``New Geometries for Gravity and Spacetime'' (IP-2018-01-7615), and also partially supported by the European Union through the European Regional Development Fund - The Competitiveness and Cohesion Operational Programme (KK.01.1.1.06).
\end{acknowledgements}


\begin{thebibliography}{}
	
	\bibitem{Hull:2001iu}
	C.~M.~Hull,
	JHEP {\bf 0109} (2001) 027
	[hep-th/0107149].
	
	\bibitem{West:2001as}
	P.~C.~West,
	Class.\ Quant.\ Grav.\  {\bf 18} (2001) 4443
	[hep-th/0104081].
	
	\bibitem{Boulanger:2003vs}
	N.~Boulanger, S.~Cnockaert and M.~Henneaux,
	JHEP {\bf 0306} (2003) 060
	[hep-th/0306023].
	
	\bibitem{Curtright:1980yk}
	T.~Curtright,
	Phys.\ Lett.\  {\bf 165B} (1985) 304.
	
\bibitem{Henneaux:2019zod}
  M.~Henneaux, V.~Lekeu and A.~Leonard,
  arXiv:1909.12706 [hep-th].
  
  \bibitem{Boulanger:2015mka}
  N.~Boulanger, P.~Sundell and P.~West,
  JHEP {\bf 1509} (2015) 192
  [arXiv:1502.07909 [hep-th]].
  
  \bibitem{Bergshoeff:2016ncb}
  E.~A.~Bergshoeff, O.~Hohm, V.~A.~Penas and F.~Riccioni,
  JHEP {\bf 1606} (2016) 026
  [arXiv:1603.07380 [hep-th]].
  
  \bibitem{Bergshoeff:2016gub}
  E.~A.~Bergshoeff, O.~Hohm and F.~Riccioni,
  Phys.\ Lett.\ B {\bf 767} (2017) 374
  [arXiv:1612.02691 [hep-th]].
  
  \bibitem{deMedeiros:2002qpr}
  P.~de Medeiros and C.~Hull,
  Commun.\ Math.\ Phys.\  {\bf 235} (2003) 255
  [hep-th/0208155].
  
  \bibitem{ChKS}
  A.~Chatzistavrakidis, G.~Karagiannis and P.~Schupp,
  arXiv:1908.11663 [hep-th].
  
  
  
  \bibitem{Chatzistavrakidis1}
  A.~Chatzistavrakidis, F.~S.~Khoo, D.~Roest and P.~Schupp,
  JHEP {\bf 1703} (2017) 070
  [arXiv:1612.05991 [hep-th]].
  
  \bibitem{Riccioni:2006az}
  F.~Riccioni and P.~C.~West,
  Phys.\ Lett.\ B {\bf 645} (2007) 286
  [hep-th/0612001].
  
  \bibitem{Boulanger:2012df}
  N.~Boulanger, P.~P.~Cook and D.~Ponomarev,
  JHEP {\bf 1209} (2012) 089
  [arXiv:1205.2277 [hep-th]].
  
  \bibitem{deBoer:2012ma}
  J.~de Boer and M.~Shigemori,
  Phys.\ Rept.\  {\bf 532} (2013) 65
  [arXiv:1209.6056 [hep-th]].
  
  \bibitem{West:2004kb}
  P.~C.~West,
  JHEP {\bf 0408} (2004) 052
  [hep-th/0406150].
  
  \bibitem{Bergshoeff:2010xc}
  E.~A.~Bergshoeff and F.~Riccioni,
  JHEP {\bf 1011} (2010) 139
  [arXiv:1009.4657 [hep-th]].
  
  \bibitem{Bergshoeff:2011se}
  E.~A.~Bergshoeff, T.~Ortin and F.~Riccioni,
  Nucl.\ Phys.\ B {\bf 856} (2012) 210
  [arXiv:1109.4484 [hep-th]].
 
 
  \bibitem{Chatzistavrakidis:2013jqa}
  A.~Chatzistavrakidis, F.~F.~Gautason, G.~Moutsopoulos and M.~Zagermann,
  Phys.\ Rev.\ D {\bf 89} (2014) no.6,  066004
  [arXiv:1309.2653 [hep-th]].
  
  \bibitem{Kimura:2014upa}
  T.~Kimura, S.~Sasaki and M.~Yata,
  JHEP {\bf 1407} (2014) 127
  [arXiv:1404.5442 [hep-th]].
  
 
  \bibitem{Tong:2002rq}
  D.~Tong,
  JHEP {\bf 0207} (2002) 013
  [hep-th/0204186].
  
  
  
  \bibitem{Becker:2009df}
  K.~Becker and S.~Sethi,
  Nucl.\ Phys.\ B {\bf 820} (2009) 1
  [arXiv:0903.3769 [hep-th]].
  
 
\bibitem{Kimura:2013zva}
T.~Kimura and S.~Sasaki,
JHEP {\bf 1308} (2013) 126
[arXiv:1305.4439 [hep-th]].

\bibitem{Bunster:2013era}
C.~Bunster and M.~Henneaux,
Phys.\ Rev.\ D {\bf 88} (2013) 085002
[arXiv:1308.2866 [hep-th]].
\end{thebibliography}
\end{document}